\documentclass[10pt]{article}
\usepackage{dcolumn}
\usepackage{bm}
\usepackage{cite}
\usepackage{amsthm}
\usepackage{amscd}
\usepackage{xspace}
\usepackage{bm}
\usepackage{amstext}
\usepackage{amsmath}
\usepackage{amsfonts}
\newcommand{\p}{\partial}

\newcommand{\dd}{{\rm d}}

\newcommand{\bd}{\begin{definition}}                
\newcommand{\ed}{\end{definition}}                  
\newcommand{\bc}{\begin{corollary}}                 
\newcommand{\ec}{\end{corollary}}                   
\newcommand{\bl}{\begin{lemma}}                     
\newcommand{\el}{\end{lemma}}                       
\newcommand{\bp}{\begin{proposition}}            
\newcommand{\ep}{\end{proposition}}                
\newcommand{\bere}{\begin{remark}}                  
\newcommand{\ere}{\end{remark}}                     

\newcommand{\bt}{\begin{theorem}}
\newcommand{\et}{\end{theorem}}

\newcommand{\be}{\begin{equation}}
\newcommand{\ee}{\end{equation}}

\newcommand{\bit}{\begin{itemize}}
\newcommand{\eit}{\end{itemize}}
\newtheorem{theorem}{Theorem}[section]
\newtheorem{corollary}[theorem]{Corollary}
\newtheorem{lemma}[theorem]{Lemma}
\newtheorem{proposition}[theorem]{Proposition}
\theoremstyle{definition}
\newtheorem{definition}[theorem]{Definition}
\theoremstyle{remark}
\newtheorem{remark}[theorem]{Remark}

\begin{document}

\title{Proper time  and conformal problem in Kaluza-Klein theory}

\author{E. Minguzzi\thanks{
Dipartimento di Matematica e Informatica ``U. Dini'', Universit\`a
degli Studi di Firenze, Via S. Marta 3,  I-50139 Firenze, Italy.
E-mail: ettore.minguzzi@unifi.it} }

\date{}
\maketitle

\begin{abstract}
In the traditional  Kaluza-Klein theory, the cylinder condition and the constancy of the extra-dimensional radius (scalar field) imply that  timelike geodesics on the 5-dimensional bundle project to solutions of the Lorentz force equation on  spacetime. This property is lost for non constant scalar fields, in fact there appear new terms that have been interpreted mainly as new forces or as due to a variable inertial mass and/or charge. Here we prove that the additional terms can be removed  if we assume that  charged particles are coupled with the same spacetime conformal structure of  neutral particles but through a different conformal factor. As a consequence, in Kaluza-Klein theory  the proper time of the charged particle might depend on the charge-to-mass ratio and the scalar field. Then we show that the compatibility between the equation of the projected geodesic and the classical limit of the Klein-Gordon equation fixes unambiguously the conformal factor of the coupling metric solving the {\em conformal ambiguity problem} of Kaluza-Klein theories. We confirm this result by explicitly constructing the projection of the Klein-Gordon equation and by showing that each Fourier mode, even for a variable scalar field, satisfies the Klein-Gordon equation on the base.
\end{abstract}


\section{Introduction}
The Kaluza-Klein (Jordan-Thiry)
theory\cite{kaluza21,klein26,lichnerowicz55} assumes the existence
of a fifth extra dimension aside the four spacetime dimensions
(for a review see
\cite{lichnerowicz55,leibowitz73,lee84,duff86,appelquist87,bailin87,collins89,overduin97,oraifeartaigh00,wesson99},
for the mathematical formalism see
\cite{geroch71,cho75,hermann78,bradfield82,hogan84,choquet-bruhat89,gladush99,betounes04}
and for generalizations see \cite{kerner68,jadczyk84,Yoon99}). Although venerable the research on this theory is still quite active \cite{bamba13,bamba14,hong14,geng14}.
This theory tried to recover gravitation and
electromagnetism on 4-dimensional spacetime $M$ from
gravitation on a 5-dimensional spacetime $P$. In
its simplest version the theory  assumes that the metric
$\tilde{g}$ on $P$ admits a spacelike Killing vector field $k$ and
that the spacetime manifold is identified with the quotient
$M=P/T_{1}$ (or $M=P/U(1)$) where $T_{1}$ (resp. $U(1)$) is  the
one-parameter group of isometries generated by $k$. In this way
$P$ is a principal bundle of projection $\pi \colon P \to M$.
In coordinate language this requirement was called {\em
cylinder condition}.
 If this condition is assumed then a number of
useful results follow, and the 5-dimensional gravitational physics
seems really to reduce to the ordinary four dimensional physics.
Clearly,   this condition would have to be justified through some physical
argument if the gravitational physics on $P$ has to be regarded as unconstrained. In this work we shall assume that the cylinder condition can
be justified in some way using an appropriate physical limit.

The Kaluza-Klein program was suggestive. It was shown that if the
scale factor of the extra dimension (scalar field) is constant in
$P$ then the geodesics of $P$ are projected to solutions of the
Lorentz force equation on $M$, where the charge-to mass ratio
depends on the geodesic chosen
\cite{leibowitz73}. Free particles on $P$ move along geodesics
thus, in this interpretation, all the particles are chargeless in
the full 5-dimensional spacetime and we see some of them as
charged only because the spacetime events are identified with the orbits of
the Killing field. However, this result was obtained assuming the
constancy of the scalar field which we denote with the letter $a$. Unfortunately,
this hypothesis in unjustified and leads to another constraint.
Indeed, it is well known \cite{overduin97} that if we assume a
constant scalar field, the Einstein equations in $P$ reduce to the
Einstein equations and the Maxwell equations in $M$ {\em plus} the
constraint that the module of the magnetic field equals the module
of the electric field, $B=E$. Thus the assumption of a constant
scalar field is in fact quite strong and the Lorentz force
equation can be recovered only if the last condition is satisfied.
The conclusion is that the Lorentz force equation is obtained only
in a very restrictive case. The situation is not really
encouraging: the constancy of $a$ implies that the geodesics
project to worldlines of charged particles but the same condition
also implies that $B=E$, a too much restrictive condition for a theory originated with unifying purposes.

A pragmatical approach to this problem, as in the original KK
theory, imposes the condition $a=cnst.$ before taking the variation of the action on $P$. In this way the Euler-Lagrange equation for $a$ is
removed {\em ad hoc}. This approach, however, not only lacks
elegance,  it is also inconsistent with the geodesic principle of
general relativity. Indeed, although not known at the time of
Kaluza, it is now well known that the geodesic principle of GR
(free particles move along  geodesics) follows from the Einstein
equations\cite[Sect.\ 9.3]{stephani82}. Therefore, the arbitrary
elimination of one of the 5-dimensional Einstein equations
questions the assumption that particles on $P$ move along
geodesics.

A better way out would be the proof that even with a non constant
scalar field the geodesics of $P$ project to solutions of the
Lorentz force equation on $M$. But this is not the case. The
projection of geodesics in the general case of non constant scalar
field has been considered in many  works
\cite{lichnerowicz55,kovacs84,gegenberg84,davidson86,cho90}. It
has been shown that new terms appear and, depending on their
arrangement, different interpretations have been proposed. Most
authors find the presence of an extra force due to the fifth
dimension and that the charge-to-mass ratio is not constant. This
last feature, which is also a departure from the standard Lorentz
force equation, raises  the question as to whether this
variability should be assigned to the mass,  the charge or both.
The same conclusions have been drawn in a version of the theory
where the cylinder condition is
dropped\cite{wesson95,wesson99,poncedeleon04}. See also the
related ``brane'' works \cite{youm00,seahra02,jalalzadeh04} (The
reader should keep in mind the distinction between recent brane
theories which are based on an embedding of the spacetime in the
higher dimensional space $E\colon M \to P$  and K-K theories which start from a
submersion $\pi\colon P \to M$, see \cite{figueiredo04}).

Moreover, its is not difficult to show that if the timelike
geodesics of $P$ project to solutions of the Lorentz force
equation then we are in the simple case of constant scalar field.
Thus there seems to be no solution to the problem mentioned
above.

Fortunately, there is a way out which  is a natural one.
We shall prove that all the extra forces that appear in the
equation of the projected geodesic have a common interpretation.
They are due to a conformal rescaling of the spacetime metric that
depends on the particle charge-to-mass ratio. With this correction to the coupling metric  we recover the Lorentz
force equation even in presence of
a non constant scalar field.
There is also a natural physical
consequence: the proper time of a charged particle in Kaluza-Klein theory depends on
its charge-to-mass ratio and on the value of the scalar field
along its path.

We shall comment more on this after the formal statement. At this
stage we wanted only to notice that  no other mechanism has been
proposed that is able  to keep the good results of the dimensional
reduction idea (the reproduction of the Lorentz force equation), while being able to remove the bad
ones (the condition $B=E$)  with the good news that there is no
need to assume a constant scalar field.

\section{Notations and terminology}
We introduce some notations and terminology.  We use units such that the speed of light $c$ satisfies the normalization $c=1$. The spacetime is a 4-dimensional
Lorentzian manifold $(M,g)$ of signature {$(+ - - -)$}. We use this signature since it is the most used in quantum field theory, and since later on we shall work with the Klein-Gordon equation.  The spacetime $M$ is the base for  a 5-dimensional bundle $\pi\colon P \to M$ where $P$ is endowed with a Lorentzian metric $\tilde g$ of signature $(+----)$. This is the usual choice since the alternative $(++---)$ does not lead to the correct relative sign between the gravitational and electromagnetic Lagrangians. The fiber of $P$  is generated by a spacelike Killing vector field $k$ on $P$.  The group structure can be the one-dimensional group of translations
$T_{1}$ or $U(1)$. With
\[
r=q/m
\]
we denote the charge-to-mass ration of a particle.

With a $\hat{} $ over a 1-form we denote the corresponding vector,
for instance $\hat\gamma^{\mu}=g^{\mu \nu}\gamma_{\nu}$.
Analogously with a $\check{}$ over a vector we denote the
corresponding 1-form, for instance $\check v_{\mu}= g_{\mu \nu}
v^{\nu}$. If the index is raised (resp.\ lowered) using the metric
$\tilde{g}_{\mu \nu}$ then we shall use the notation
$\hat{\tilde{\gamma}}$ (resp. $\check{\tilde{v}}$). The 1-form on
$P$
\begin{equation}
\tilde{\omega}=\check{\tilde{k}}/\tilde{g}(k, k)
\end{equation}
is the real-valued connection 1-form \cite{kobayashi63}. It satisfies
$\tilde{\omega}(k)=1$ and $L_{k} \tilde{\omega}=0$ where $L_{k}$
is the Lie derivative. Let $U$ be an open subset of $M$ and let $s\colon U \to P$ be a section: $\pi\circ s=Id_M$.  Defined the electromagnetic potential as the pullback $A=s^*\tilde\omega/\beta$, where $\beta$ is a dimensional constant, we have that
the connection 1-form can be written
\begin{equation}
\tilde{\omega}=\dd y+ \beta A_{\mu} \dd x^{\mu}
\end{equation}
where $y$ is a dimensionless coordinate on the fibre and
$\{x^{\mu}\}$ are coordinates on $U$.
The electromagnetic
field $F:=\dd A$ is independent of the section and can be regarded
as the pullback under a local section of $\dd
\tilde{\omega}/\beta$. With $\hat{F}$ we denote the
electromagnetic field with the first index raised,
$\hat{F}^{\mu}_{\ \ \nu}=g^{\mu \alpha} F_{\alpha \nu}$. In what
follows we shall be involved also with metrics denoted $g_{E \,
\mu \nu}$ and $g_{r\, \mu \nu}$. We shall therefore write
$\hat{F}^{ \ \mu}_{E \ \nu}=g^{\mu \alpha}_{E} F_{\alpha \nu}$ and
$\hat{F}_{r}{}^{ \mu}_{\ \nu}=g^{\mu \alpha}_{r} F_{\alpha
\nu}$. This index free notation will shorten the formulas throughout  the work.
The positive scalar field $a$ is defined by
\[a^{2}=-\tilde{g}(k,k).\]

It remains to define a Lorentzian metric on  $M$. There are two
usual choices: the Jordan metric
\begin{equation} \label{jordan}
g_0=\tilde{g}-\frac{\check{\tilde{k}} \otimes
\check{\tilde{k}}}{\tilde{g}(k,k)}=\tilde{g}+a^{2} \tilde{\omega}^{2},
\end{equation}
and the Einstein metric ($a_{0}$ is a constant with the same
dimension of $a$.)
\begin{equation} \label{eframe}
g_{E}=(a/a_{0})g_0.
\end{equation}
Actually, here we are really giving the representative of the spacetime metric on $P$. The metric can be passed  to the quotient to $M$ since $g(k,\cdot)=0$ and $L_kg=0$, see \cite{geroch71}.

The freedom in the choice of spacetime metric is due to the arbitrariness in the choice of the
conformal factor in front of the spacetime metric $g$ and this
complication  is sometimes called the {\em conformal
ambiguity problem}
\cite{sokolowski89,ferraris90,damour92,capozziello97,faraoni99,flanagan04}.
It is not present in the constant scalar field case since there
the two metrics differ by a constant factor with no relevant
physical consequences. As we shall see our results on the Lorentz
force equation will, in some sense, privilege the Jordan metric
for atomic phenomena. This does not mean that the Jordan metric is
the ``right'' spacetime metric. It has become clear
\cite{fierz56,flanagan04} that in gravitational theories with
scalar fields there is no unique spacetime metric. The metric with
which one should calculate the proper time of a particle can
differ for different kinds of particles if the corresponding matter
Lagrangian terms couple to different metrics. In particular the
metric that appears in the free electron Lagrangian may be only
conformally related to the Einstein metric i.e. to the metric that
appears in the Einstein-Hilbert Lagrangian term.
\begin{align*} \label{action}
-\int \dd y \,\dd^{4} x\sqrt{\vert \textrm{det} \tilde{g}
\vert}\tilde{R}&= -(\int\!\! \dd y)\int \dd^{4}x
\sqrt{\vert \textrm{det} {g}_{E} \vert}\,
 \big\{a_{0} R_{E} \nonumber \\
 &\quad+\frac{a^{3}\beta^{2}}{4} \hat{F}_{E}^{2} -\frac{3a_{0}}{2} g^{\mu \nu}_{E}(\p_{\mu} \ln a)  (\p_{\nu} \ln a)  \big\}
\end{align*}
(a total divergence term $D=-a_{0} \, \Box_{E} \ln a$ has
been integrated out).

Thus, the
number of rotations that a binary black hole system
performs would be proportional to the proper time calculated with
the cosmological Einstein metric, while the proper time of a
charged particle in the same cosmological background would have to be
calculated with a metric that is only conformally related to
Einstein's. As we shall see the present work will  fix that
conformal factor providing a solution to the conformal ambiguity
problem.

The constant $a_{0}$ appearing in (\ref{eframe}) is the present
value of $a$ (if $a$ changes only over cosmological scales it is
the value at this cosmological era, if it changes over the solar
system it is the value at  the earth's surface). The reason is
that the observers independently of whether they measure time with
atomic clocks, pendulums or the planets motion set their clocks in
such a way that they have the same rate {\em here and now}
(syntonization process). Since $a$ varies, after an interval that
may be huge the clocks of different nature, atomic and gravitational,
 finally desyntonize. The statement that they are set in
such a way that they measure the same unit of time in mathematical
terms is $g=g_{E}$ here and now, or $a/a_{0}(\textrm{here and
now})=1$. The interpretation of $a_{0}$ follows.

Finally, note that the equivalence principle is satisfied only if
the metric that couples to matter is the same for every kind of
massive particle\cite{fierz56}. Of course slight differences in
the coupling metrics are possible, if the experimental precision
with which the equivalence principle has been tested is taken into
account. In our case the coupling metric will depend on the
particle charge and will reduce to the Jordan metric in the
neutral case. We are therefore interested in tests  of the
equivalence principle for charged particles. Only one
experiment was dedicated to this question so far: the Witteborn-Fairbank
experiment \cite{witteborn67}. The test confirmed the equivalence
principle but the accuracy was very poor, about $0.1$ which should
be contrasted with $10^{-12}$ for neutral matter \cite{dittus04}.
More refined experiments have been proposed \cite{dittus04}.

\section{Proper time of charged particles}
Equation (\ref{jordan}) implies that the projection of a timelike (causal) curve on $P$ is timelike (resp.\ causal) on $M$. We are interested on the projection of geodesics $z: I \to P$,
\[
\tilde \nabla_t z_t=0 ,
\]
where $\tilde \nabla$ is the Levi-Civita connection of $\tilde g$. We begin with a theorem.
\begin{theorem} \label{te1}
Let $z: I \to P$, $t\mapsto z(t)$, be a geodesic, $\tilde{g}(\dot z,\dot z)=\epsilon$, $\epsilon=-1,0,1$, and let $x=\pi\circ z$ be its projection to $M$, then
\begin{equation} \label{qoverm}
r=-\beta\, {a^{2}(x(t))\,\tilde{\omega}(\dot z)}=\beta \tilde{g}(k,\dot z) ,
\end{equation}
is a constant of motion. Define on $M$ the function
\begin{equation} \label{qsum}
\Omega_{r}=[\epsilon+{r^2}/({\beta^{2}a^{2}})]^{1/2} ,
\end{equation}
which for $\epsilon<0$ we assume to be positive, and  define the Lorentzian metric
\begin{equation} \label{gqsum}
g_{r}=\Omega^{2}_{r}(\tilde{g}+a^{2}\tilde{\omega}^{2}) ,
\end{equation}
then the projection $x$ is timelike and once
parametrized with
\begin{equation} \label{llp}
t_{r}=\int \Omega^{2}_{r}(t) \,\dd
t ,
\end{equation}
it satisfies the Lorentz force
equation
\begin{equation} \label{good}
 \nabla^{(r)}_{t_r} x_{t_r}=r \hat{F}_{r}(x_{t_r}), \qquad
g_{r}(x_{t_r},x_{t_r})=1 ,
\end{equation}
where $\nabla^{(r)}$ denotes the Levi-Civita  connection of $g_{r}$ and $\hat F_r{}^{\mu}_{\ \nu}=g_{r}^{\mu \alpha}F_{\alpha \nu}$.\\
\end{theorem}

The meaning of the theorem is that a charged particle {\em sees}
the same causal structure of neutral particles but a different
conformal factor. That factor depends on the charge-to-mass ratio
of the particle and is close to $1$ for great values of the scalar
field $a$. The parametrization $t_{r}$ is the
$r$-proper time of the particle i.e.\ that obtained from
the metric $g_{r}$ integrating the line element over
$x$. The theorem suggests that for charged particles this
proper time is more fundamental than the usual $0$-proper time
$t_0$, obtained integrating the Jordan metric
over $x$.

We can think of an elementary particle as a clock. Indeed,
elementary particles can decay and this decay time has been
measured in different reference frames to give an experimental
verification of the time dilation predicted by special relativity.
The previous theorem suggests that the time of a charged particle
clock is given by $t_{r}$. This clock is therefore
predicted to go faster for small $a$ indeed\footnote{The parameters $t_{0}$ and $t$ should not be confused, the former is the proper time according to the Jordan metric on $M$, the latter is the proper time parametrization of the geodesic on $P$. They
differ along the trajectory of a charged particle, from (\ref{gqsum}) we have $\dd
t_{r}=\Omega_{r} \dd
t_{0}$ while from (\ref{llp}) we have $\dd
t_{r}= \Omega_{r}^{2} \dd t $.} for the classical case in which the geodesic on $P$ is timelike ($\epsilon =1$)
\begin{equation}
\dd
t_{r}=\sqrt{1+\frac{1}{\beta^{2}a^{2}}(\frac{q}{m})^{2}}
\, \dd t_{0}.
\end{equation}


Since the conformal factor depends on the charge-to-mass
ratio we have violation of the equivalence principle which is weak for a slowly varying
scalar field. Indeed, we are not
saying that the usual extra force is not present but only that it
is an apparent force due to an unnatural metric choice. The
analogy with the generalization from the Newtonian to the
Einsteinian gravitation is clear: the Newton force arises only in
a formalism that unnaturally forces the metric to be the flat one.
In the right formalism the motion appears unforced  and one also
realizes that the spacetime metric of the new theory is related to
 proper time. Exactly the same happens here. We remove the
extra force and, as a consequence, we find a formula for the
proper time of charged particles.

Let us give a proof Theorem \ref{te1}. A related variational observation was given by Lichernowicz and Thiry
\cite{lichnerowicz55} as a particular case of a general result on
the dimensional reduction of Finsler geometries.
\begin{proof}
We start from the local expression of the metric
\begin{equation} \label{moc}
\tilde{g}=\Omega^{-2} g-a^{2}(\dd y+ \beta A)^{2} ,
\end{equation}
where $\Omega$, $g$ and $A$ are tensors on $U_{i}$. Let $z$ be a curve on $P$ and let $x=\pi\circ z$ be its projection, we have
\begin{equation} \label{sq}
\tilde{g}(\dot z,\dot z)=\Omega^{-2}g(\dot x,\dot x)-a^{2}(\dot
y+\beta A(\dot x))^{2} .
\end{equation}
Thus using a variational trick in $\pi^{-1}(U_{i})$
\begin{align*}
0&=\delta\frac{1}{2} \int \dd t\, \big\{
\tilde{g}(\dot z,\dot z)-\Omega^{-2}g(\dot x,\dot x)+a^{2}(\dot
y+\beta A(\dot x))^{2}\big\} \\
&=\int \dd t\, \big\{ -\tilde{g}(\tilde{\nabla}_t \dot z,\delta
z)+\Omega^{-2}g(\nabla_t \dot x,\delta x)-(\frac{\delta\Omega^{-2}}{2})g(\dot x,\dot x)\\
&\quad +\frac{\dd \Omega^{-2}}{\dd
t}\,g(\dot x,\delta x) +(\frac{\delta
a^2}{2})(\dot y +\beta A(\dot x))^{2} +a^{2}(\dot y+\beta A(\dot x))(\dot{\delta
y}+\beta \delta(A(\dot x)))\big\} \\
&=\int \dd t\,  \big\{ -\tilde{g}(\tilde{\nabla}_t \dot z,\delta
z)+\Omega^{-2}g(\nabla_t \dot x,\delta x)-(\frac{\delta\Omega^{-2}}{2})g(\dot x,\dot x)\\
&\quad +\frac{\dd \Omega^{-2}}{\dd
t}\,g(\dot x,\delta x) -\frac{1}{2}(\delta
\frac{1}{a^{2}}) [a^{2}\tilde{\omega}(\dot z)]^{2}-\tilde{\omega}(\delta z)\frac{\dd}{\dd
t}[a^{2}\tilde{\omega}(\dot z)]+\beta [
a^{2}\tilde{\omega}(\dot z)]F(\delta x,\dot x)\big\}
\end{align*}
where $\tilde{\omega}(\dot z)=\dot y+ \beta
A(\dot x)$, $\tilde{\omega}(\delta z)=\delta y+ \beta
A(\delta x)$ and where $\tilde \nabla$ and $\nabla$ are the Levi-Civita connections of $\tilde g$ and $g$ respectively. Note that
\begin{equation}
\delta \Omega^{-2}= \delta x^{\nu} \p_{\nu}
\Omega^{2}=g(\delta x, \nabla \Omega^{-2})
\end{equation}
and analogously for $\delta(1/a^{2})$. The vertical variation
gives ($\delta x=0$)
\begin{equation} \label{first}
a^{2}\tilde{\omega}(\tilde{\nabla}_t \dot z)=\frac{\dd}{\dd
t}[a^{2}\tilde{\omega}(\dot z)],
\end{equation}
while the horizontal variation ($\tilde{\omega}(\delta z)=0$)
gives
\begin{align}
\begin{split}
\pi_{*}\tilde{\nabla}_t \dot z&={\nabla}_t \dot x-\frac{\dd\ln
\Omega^{2}}{\dd
t}\,\dot x+\beta {[a^{2}\tilde{\omega}(\dot z)]} \Omega^{2} \hat{F}(\dot x) \\
&\quad -\frac{\Omega^{2}}{2}\big\{(\nabla\Omega^{-2})g(\dot x,\dot x)+
(\nabla\frac{1}{a^{2}})[a^{2}\tilde{\omega}(\dot z)]^{2}\big\}. \label{second}
\end{split}
\end{align}
Now, let $z(t)$ be a  geodesic of $P$ with the
affine parametrization $t$ such that
$\tilde{g}(\dot z,\dot z)=\epsilon=1,0,-1$. Eq.\ (\ref{first})
states that the quantity $a^{2}\tilde{\omega}(\dot z)$ is a
constant of motion and defining $r=q/m=-\beta
a^{2}\tilde{\omega}(\dot z)$,  Eq.\ (\ref{second}) which
determines the trajectory of the projected geodesic reads
\[
\nabla_t \dot x-\frac{\dd\ln \Omega^{2}}{\dd
t}\, \dot x=\Omega^{2}\,r\hat{F}(\dot x) +\frac{\Omega^{2}}{2}\big\{(\nabla\Omega^{-2})g(\dot x,\dot x)+
(\nabla\frac{r^{2}}{\beta^{2}a^{2}})\big\}.
\]
Let us reparametrize the curve $x$ with respect to  a parameter
$t'$ such that $\dd t'=\Omega^{2}\dd t$ and use
\begin{equation}
\nabla_{t'} x'=\Omega^{-4}(\nabla_t \dot x-\frac{\dd\ln \Omega^{2}}{\dd
t}\, \dot x)
\end{equation}
with Eq.\ (\ref{sq}) we obtain
\begin{align*}
\nabla_{t'} x'&= r \hat{F}(x')
+\frac{1}{2}\{(\nabla\Omega^{-2})[\epsilon+\frac{r^{2}}{\beta^{2}a^{2}} ]
+
\Omega^{-2}(\nabla\frac{r^{2}}{\beta^{2}a^{2}})\} \nonumber \\
&= r\hat{F}(x')
+\frac{1}{2}\nabla\{\Omega^{-2}[\epsilon+\frac{r^{2}}{\beta^{2}a^{2}} ]\}.
\end{align*}
Thus with the choice $\Omega^{2}=C  [
\epsilon+\frac{r^{2}}{\beta^{2}a^{2}}  ]$, $C=cnst$, we obtain the Lorentz force equation and hence the fact that  $g(x',x')$ is constant. From Eq.\ (\ref{moc}) using $\dot x=\Omega^2x'$ we obtain $g(x',x')=C^{-1}$. This concludes the proof, however we wish to  observe that the found formulas  for
$\epsilon=-1$, $C=1$,  imply the interesting result:
{\em if the charged particle is the projection of a tachyon in $P$ then its
motion is confined to the region $a(x) < \frac{q}{\beta m}$}.
\end{proof}

\section{The conformal ambiguity problem}
Let us investigate the conformal ambiguity problem in the more general framework of field theories. Let us start again
with a  spacetime $(M,g)$ and let us consider the Klein-Gordon
Lagrangian
\begin{equation}
\mathcal{L}_{\psi}=\sqrt{\vert \det g \vert }\,\big\{g^{\mu \nu}(D^{q}_{\mu}\psi)^{*}D^{q}_{\nu}
\psi-\epsilon m^{2}\psi^{*}\psi \big\},
\end{equation}
where $D^{q}_{\mu}=\hbar \nabla_{\mu}-i q A_{\mu}$, and $\epsilon=-1,0,1$, where $\epsilon =1$ is the usual case.\footnote{We shall  also use this Lagrangian on $P$ where we have seen that spacelike, lightlike or timelike geodesics might project to timelike solutions of the Lorentz force equation. This fact signals that it is equally interesting to  study the K-G equation for lightlike or tachyonic matter on $P$.}

The field
$\psi$  which represents a charged spinless particle satisfies the
Klein-Gordon equation
\begin{equation}
D^{q\, \mu}D^{q}_{\mu}\psi+\epsilon m^{2}\psi=0 .
\end{equation}
Now we are going to perform a short-wave (WKB) approximation to
this equation in order to recover the classical limit. We expand
$\psi  $ in powers of $\hbar$
\begin{equation}
\psi=(\psi_{0}+\hbar\psi_{1}+
\mathcal{O}(\hbar^{2}))\, e^{\frac{i}{\hbar}S} ,
\end{equation}
where $\hbar^{-1}S$ is the phase. Plugging this expression into the
Klein-Gordon equation and equating to zero the terms of order
$\hbar^{0}$ we obtain
\begin{equation} \label{hj}
(\nabla^{\mu}S-qA^{\mu})(\nabla_{\mu}S-qA_{\mu})-\epsilon m^{2}=0 .
\end{equation}
This equation can be regarded as the Hamilton-Jacobi equation
associated to the (super)Hamiltonian
$\mathcal{H}=(P^{\mu}-qA^{\mu})(P_{\mu}-qA_{\mu})-\epsilon m^{2}=0$, where
$P_{\mu}=S_{;\mu}$ is the conjugate momentum. We define the
kinematical momentum $p_{\mu}:=S_{;\mu}-qA_{\mu}$ and note that
\begin{equation}
p_{\alpha;\beta}=S_{;\alpha;\beta}-qA_{\alpha;\beta}=p_{\beta;\alpha}+qF_{\alpha
\beta}.
\end{equation}
We observe that
\begin{align}
\begin{split}\label{sis1}
p^{\alpha}_{\,;\beta}p^{\beta}&=g^{\alpha
\gamma}p_{\gamma;\beta}p^{\beta}=g^{\alpha
\gamma}(p_{\beta;\gamma}+qF_{\gamma \beta})p^{\beta} \\
&= \frac{1}{2}g^{\alpha \gamma}(p^{\mu}p_{\mu})_{;\gamma}+q
F^{\alpha}_{\ \ \beta}p^{\beta}=q F^{\alpha}_{\ \ \beta}p^{\beta} ,
\end{split}
\end{align}
where we have used Eq. (\ref{hj})
\begin{equation} \label{sis2}
p^{\mu}p_{\mu}=\epsilon m^{2} .
\end{equation}
Let $\epsilon=1$. Equations (\ref{sis1})-(\ref{sis2}) form a system which states
that the integral lines $x(s)$ of the kinematical momentum $\frac{\dd x^{\mu}}{\dd s
}=p^{\mu}/m$ are solutions to the Lorentz force equation relative
to the charge-to-mass ratio $q/m$. As $p^{\mu}/m$ is
normalized, $s$ is the proper time parametrization along the
integral curve $x$. In other words the classical limit of the
Klein-Gordon equation gives, as expected, the Lorentz force
equation.
Over each wordline $x(s)$ there is a
natural time given by the number of wavefronts of $\psi$ seen by
an observer moving along $x(s)$, that is the phase $S$ provides a natural clock over $x(s)$. However, here there is a difficulty
connected to the fact that the Klein-Gordon Lagrangian is
invariant under gauge transformations, thus $S$ is not determined
by the initial conditions since it is not gauge invariant. Instead
the 1-form $dS-qA$ is gauge invariant. Its integral over $x$
provides the natural clock we were looking for. It can be
interpreted as the number of wavefronts of $\psi$ seen by an
observer moving along $x(s)$. If the gauge has been fixed so that
$A_{\mu}(x(s))\frac{\dd x^{\mu}}{\dd s}=0$
\begin{equation}
\frac{\dd S}{\dd s}=\frac{\dd S}{\dd s}-qA_{\mu} \frac{\dd x^{\mu}}{\dd s}
=p_{\mu}{p^{\mu}}/{m}=\epsilon m
\end{equation}
thus the natural clock given by the gauge fixed phase $S$
coincides, suitably normalized, and for $\epsilon=1$ with the usual proper time
parametrization derived from the metric $\dd s^{2}=g_{\mu \nu}\dd
x^{\mu} \dd x^{\nu}$.

Let us now come to the Kaluza-Klein case. If  a charged particle
is represented by a geodesic on $P$ then its projection satisfies
the Lorentz force equation on $M$ with respect to the metric
$g_{q/m}$. Then the classical 4-dimensional Lagrangian for the
charged particle must be
\begin{equation}
\mathcal{L}_{\psi}=\sqrt{\vert\det g_r\vert}\, \big\{g_{r}^{\mu \nu}(D^{(q,m)}_{\mu}\psi)^{*}D^{(q,m)}_{\nu}
\psi-m^{2}\psi^{*}\psi\big\}
\end{equation}
where \[D^{(q,m)}_\mu=\hbar\nabla^{(r)}_\mu-iq A_\mu\]
that is, it should couple with $g_{r}$ and not with $g_{0}$,
otherwise the
classical limit of the Klein-Gordon equation would give a particle
motion that cannot be derived from a geodesic projection. This
argument solves the conformal ambiguity problem of Kaluza-Klein
theory since it fixes the conformal factor in the coupling metric.
The previous analysis also tells us that the most natural proper
time over the classical limit trajectories, i.e.\ the gauge fixed
field phase, is nothing but the proper time calculated with the
metric that appears in the Lagrangian, that is $\dd
t_r^{2}=g_{r \, \mu \nu} \dd x^{\mu} \dd x^{\nu}$. This fact
confirms again the interpretation of $t_r$  as proper time of the charged-particle.

Finally, note that since $\psi$ couples with $g_{r}$ and no term
in $\psi$ couples with $a$, and since both $g_{0}$ and $g_{E}$ can
be recovered only from the knowledge of {\em both} $g_{r}$ and
$a$ it is simply impossible that these last metrics have some role
in the proper time of the charged particle represented by $\psi$.
The argument is: if the proper time of the particle depends on
$g_{0}$ ($g_{E}$) then $\psi$ should have a dynamics which depends
on $a$ too (and not solely on $g_{r}$), but this is in
contradiction with the Lagrangian expression.

In the next section we shall see that the coupling metric needs a
correction and that it is given by $g_{r}$ only in the limit
$\hbar \to 0$.

\section{Conformal problem and quantum mechanics} \label{ca2}

Let us consider the Klein-Gordon equation for neutral particles  in $P$ (here $A,B=0,1,\cdots 4$)
\begin{equation} \label{kos}
\hbar^{2}\tilde{g}^{A B}\tilde{\nabla}_{A} \tilde{\nabla}_{B} \Psi
-\epsilon \,{m}^{2} \Psi=0, \quad \epsilon=1.
\end{equation}
For more generality we have introduced a parameter $\epsilon$ which takes values in $\{-1,0,1\}$ and whose purpose will be explained in a moment.

From the results of the previous section it follows that this equation
returns the geodesic equation on $P$ under the classical  short wave approximation  ($\hbar \to 0$).
 That is, if $\Psi =|\Psi|
e^{\frac{i}{\hbar}\tilde{S}}$ with $|\Psi|=\Psi_{0}+\Psi_{1} \hbar
+ O(\hbar^{2})$ then the vector field $u_{A}=\p_{A}\tilde{S}/{m}$
is normalized, $\tilde g(u,u)=\epsilon$, and $\tilde{\nabla}_{u}u=0$. The momentum of the
particle is $\tilde{p}_{A}={m} u_{A} = \p_{A}\tilde{S}$. In this
short wave limit
$\tilde{p}_{y}=\p_{y}\tilde{S}=k^{A}\tilde{p}_{A}$ is constant
along the integral geodesics since
$\tilde{\nabla}_u\tilde{g}(k,u)=0$.

This quantity although
constant along a given geodesic may vary changing the integral
geodesic in $P$. Using the expression for $\tilde{g}$, $k=\p_y$ and Eq.\ (\ref{qsum}) it is easily shown that $q=\beta\tilde{p}_{y}$. We know that these geodesics projects on solutions of the Lorentz
force equation with respect to $g_{r}$, $r=q/m$. But we also know that
the Klein-Gordon equation  in $M$ of parameters $q$ and $m$ and
coupled to $g_{r}$ gives in the classical limit the same Lorentz
force equation. In general one expects that the Klein-Gordon
equation for a metric $g_{(q,m)}$ such that $g_{(q,m)} \to
g_{r}$ in the classical limit,  would have as a classical limit
the same Lorentz force equation. Thus maybe there is a choice of
$g_{(q,m)}$ and a projection of the Klein-Gordon equation such
that the following diagram commutes
\[\begin{CD}
 \textrm{neutral KGE} \ \textrm{in} \ (P, \tilde{g}) @>\hbar \to 0>> \textrm{GE} \ \textrm{in} \ (P, \tilde{g}) \\
@V?V{\pi}V  @VV{\pi}V \\
\textrm{charged KGE} \ \textrm{in} \ (M, g_{(q,m)}) @>\hbar \to 0>>
\textrm{LFE} \ \textrm{in} \ (M, g_{q/m})
\end{CD}\]
where GE stands for {\em geodesic equation} and similarly for KGE and LFE.
The problem is thus whether the K-G equation in $P$ can be reduced
to the K-G equation in $M$ coupled to a metric $g_{(q,m)}$ before
the classical limit is taken. If this is possible then the
classical limit can be performed irrespective of the order both in $P$ and $M$.

There is a difficulty, however. In the classical limit,
 the K-G equation on $M$ determines a
solution of the Lorentz force equation of charge-to-mass ratio
$q/m$ where both $q$ and $m$ are fixed by the variables entering the K-G equation.
On the contrary, there is no variable $q$ in the K-G equation on $P$, we have just a constant $q$ given by $\beta \p_{y}\tilde{S}$ over the geodesics on $P$ obtained with the short wave approximation.
The K-G equation in $P$ can project to the K-G equation in $M$ only if
the classical limit of the K-G equation in $P$ projects to the
classical limit of the K-G equation in $M$. This is possible only
if  $q$ is a constant  all over $P$, which means
$\p_{y}\tilde{S}=q/\beta$.

Consider the Fourier expansion along the fiber direction
\begin{equation}
\Psi=\int d\mu(q)\, \bar{\psi}_{q}(x) \,e^{i\frac{q}{\hbar\beta} y}
\end{equation}
where $d \mu(q)$ is a suitable measure; that is $\int d\mu(q)=\int
\dd q/(2\pi)$ if the fiber is $\mathbb{R}$ or $\int
d\mu(q)=\Sigma_{n}$ with $n=q/\hbar\beta$ otherwise. If the
extradimension is compactified this is the usual Fourier series
and hence $q=ne$ where $e=\hbar\beta$ is a fundamental charge. Now
note that each term of the series satisfies the  above condition $\p_{y}\tilde{S}=q/\beta$.
The problem we want to solve is therefore:
\begin{quote}
{ Given $\Psi=\int d \mu(q) \, \bar{\psi}_{q}(x)
e^{i\frac{q}{\hbar\beta} y}$  is it possible to reduce the K-G
equation in $P$ to a 1-parameter ($q$) family of K-G equations in
$M$?}
\end{quote}
This is an old problem. It was solved affirmatively by Klein for a constant scalar field $a$. We are going to provide an
affirmative solution  for the non constant scalar field
case. Actually, we shall even prove that the answer is affirmative  for any value  $\epsilon=-1,0,1$ and so also for $\epsilon=0$ which means that the Klein-Gordon field can describe massless matter on $P$. This fact  is consistent with the possibility of obtaining the Lorentz force equation on $M$ from geodesics of any causal type on $P$.
As expected, however, the metric on the base will change with
$q$ and in the classical limit will be $g_{q/m}$.

Let us write
\begin{equation}
\tilde{g}=\Omega^{-2} g-a^{2} (\dd y + \beta A_{\mu} \dd
x^{\mu})^{2}
\end{equation}
where the function $\Omega$ (and hence the metric $g$) has to be
determined.

\begin{remark}
Actually, the next calculation will hold for $\Omega^2<0$ as well, which means that $\tilde{g}$ might have signature $(-,-,+,+,+)$ where the fiber of $P$ is timelike. We denote $\varepsilon=\textrm{sign}(\Omega^2)$, so the next expressions will depend on $(\epsilon,\varepsilon)$ the standard Kaluza-Klein theory corresponding to $(1,1)$.
\end{remark}

 The vectors $e_{\mu}=\p_{\mu}-\beta A_{\mu} \p_{y}$
and $k=\p_{y}$ give the dual base to $\{\dd x^\mu, \tilde \omega\}$ and satisfy
\begin{align*}
\tilde{g}(k,k)&=-a^{2}, \quad
\tilde{g}(k,e_{\mu})=0, \quad
\tilde{g}(e_{\mu},e_{\nu})= \Omega^{-2} g_{\mu \nu},
\end{align*}
so the inverse metric on $P$ is
$\tilde{g}^{-1}=\Omega^{2}g^{\mu \nu}e_{\mu}\otimes e_{\nu}-a^{-2}\, \p_y\otimes \p_y$.
The K-G Lagrangian can be rewritten
\begin{align*}
\mathcal{L}_{\Psi}&=\sqrt{\vert\textrm{det}\tilde{g}\vert}\, (\hbar^{2}\tilde{g}^{A
B}{\p}_{A} \Psi^{*}{\p}_{B} \Psi -\epsilon {m}^{2}
\Psi^{*}\Psi)\\
&=\sqrt{\vert\textrm{det}g\vert}\,
\Omega^{-4}a\big\{\hbar^{2}\Omega^{2} g^{\mu
\nu}e_{\mu}[\Psi^{*}] e_{\nu}[\Psi]-\frac{\hbar^{2}}{a^{2}}\vert
\p_{y}\Psi\vert^{2} -\epsilon {m}^{2} \Psi^{*}\Psi \big\}
\end{align*}
Taking the variation with respect to $\Psi$ and ignoring total divergence terms with respect to the measure $\dd^4 x\dd y$  we find the K-G
equation
\begin{equation*}
-\Omega^{-2}a g^{\mu \nu} (\nabla_{\mu}-\beta A_{\mu} \p_{y})
e_{\nu}[\Psi]-g^{\mu
\nu}\p_{\mu}(\Omega^{-2}a)e_{\nu}[\Psi]=-\frac{\Omega^{-4}}{a}\,
\p^{2}_{y}\Psi+\frac{\epsilon m^{2}}{\hbar^{2}}\,\Omega^{-4}a\Psi .
\end{equation*}
In practice we have used the variational formulation of the K-G
equation as a tool in order to rewrite the K-G equation in terms of the Levi-Civita covariant
derivative of $g$.
The previous equation can be rewritten
\begin{align*}
g^{\mu \nu}&\big(\nabla_{\mu}-\beta
A_{\mu}
\p_{y}-\frac{1}{2}\,\p_{\mu}\ln\vert \Omega^{2}/a\vert\big)\,\big(\nabla_{\nu}-\beta
A_{\nu}
\p_{y}-\frac{1}{2}\,\p_{\nu}\ln\vert \Omega^{2}/a\vert\big)\Psi\\
&=\big\{-\frac{1}{2}\,\Box\ln\vert \Omega^{2}/a\vert
+\frac{1}{4}\,\p_{\mu}\ln\vert\Omega^{2}/a\vert\,\p^{\mu}\ln\vert\Omega^{2}/a\vert
-\frac{\epsilon m^{2}}{\hbar^{2}}\,\Omega^{-2}\big\}\Psi+\frac{\Omega^{-2}}{a^{2}}\,\p^{2}_{y}
\Psi .
\end{align*}
Plugging the Fourier expansion into this equation we obtain for
each $q$
\begin{align*}
&\big(D^{q \,
\mu}-\frac{\hbar}{2}\,\p^{\mu}\ln\vert\Omega^{2}/a\vert\big)\big(D^{q}_{\mu}-\frac{\hbar}{2}\,\p_{\mu}\ln\vert\Omega^{2}/a\vert\big)\bar{\psi}_{q}\\
&=\big\{-\frac{\hbar^{2}}{2}\,\Box\ln\vert\Omega^{2}/a\vert
+\frac{\hbar^{2}}{4}\, \p_{\mu}\ln\vert\Omega^{2}/a\vert\p^{\mu}\ln\vert\Omega^{2}/a\vert
-\epsilon m^{2}\Omega^{-2}-\frac{\Omega^{-2} q^{2}}{\beta^{2}
a^{2}}\big\} \bar{\psi}_{q} .
\end{align*}
Now define $\psi_{q}$ such that
$\bar{\psi}_{q}=\psi_{q}\,m\vert\Omega^2 a_0/a\vert^{1/2}$ 
and choose $\Omega \equiv \Omega_{(q,m)}$ where $\Omega_{(q,m)}$
satisfies the differential equation
\begin{align}
\begin{split}\label{omegadif}
-\frac{\hbar^{2}}{2}\,\Box\ln\vert \Omega_{(q,m)}^{2}/a\vert
&+\frac{\hbar^{2}}{4}\,g^{\mu \nu}\p_{\mu}\ln\vert \Omega_{(q,m)}^{2}/a\vert\,\p_{\nu}\ln\vert\Omega_{(q,m)}^{2}/a\vert
\\
&-\Omega_{(q,m)}^{-2}[\epsilon m^{2}+\frac{q^{2}}{\beta^{2}
a^{2}}]=-m^{2} ,
\end{split}
\end{align}
then $\psi_{q}$ satisfies the K-G equation
\begin{equation}
D^{(q,m) \, \mu}D^{(q,m)}_{\mu}\psi_{q}+m^{2}\psi_{q}=0 ,
\end{equation}
with respect to the metric $g_{(q,m)}=\Omega_{(q,m)}^{2}g_{0}$.
Here the conclusion is the same as in the study of the Lorentz
force equation. There is a metric that allows us to write the
usual K-G equation, however this metric is only conformally related to
$g_{0}$.

We see that this analysis is, as expected, consistent with that
for the Lorentz force equation. Indeed, taking the classical limit
$\hbar \to 0$ we find from (\ref{omegadif}), $\Omega_{(q,m)} \to
\Omega_{q/m}$ and hence $g_{(q,m)} \to g_{q/m}$ in this limit.
This section generalizes therefore the result of the previous one.
Here the conformal ambiguity problem is solved also for
$\hbar \ne 0$.

 Another case in which we obtain the classical solution $\Omega_{(q,m)} =
\Omega_{q/m}$  is that in which $a$ is constant, indeed with $\Omega_{(q,m)}=cnst$ the first two terms in (\ref{omegadif}) vanish. Of course, even if $a$ is constant Eq.ù (\ref{omegadif}) need not have only this solution for $\Omega_{(q,m)}$, since the general solution will depend on the initial conditions.

The solution  is not available in closed
form as a function of $a$. Instead one has to solve a differential
equation. Note that Eq.\ (\ref{omegadif}) cannot be solved as it
stands. In fact there appear covariant derivatives of $g_{(q,m)}$
while $g_{(q,m)}$ is known only when $\Omega_{(q,m)}$ is known. It
is more convenient to rewrite it in terms of
$g_{E}= \vert\Omega_{(q,m)}^{-2}\, \frac{a}{a_{0}}\vert\,  g_{(q,m)}$ and its
covariant derivative. Then the Einstein equations in 5-dimensions
give $g_{E}$, $A$ and $a$, while $\Omega_{(q,m)}$ (and hence
$g_{(q,m)}$) will be determined solving another differential
equation equivalent to (\ref{omegadif}). In order to find this
equation let us note that, for any function $f$ if
$g_{E}=\vert  \Omega^{-2}\, \frac{a}{a_{0}}\vert  g$ we have (see e.g.\ \cite[Eq.\ (D.12)]{wald84})
\begin{equation}
\Box f=\vert\Omega^{-2}\frac{a}{a_{0}}\vert\big(\Box_{E}f
+g_{E}^{\mu \nu}\p_{\mu}\ln\vert\Omega^{2}/a\vert \, \p_{\nu}f\big) ,
\end{equation}
and defining
$\alpha_{(q,m)}^2:=m^2\vert\Omega_{(q,m)}^2\frac{a_{0}}{a}\vert$ and recalling that $\varepsilon=\textrm{sign} \,\Omega^2$
 we find
\begin{equation}
\hbar^{2}\Box_{E}\,\alpha_{(q,m)}+ \varepsilon\frac{a_{0}}{a}\, \big[\epsilon m^2+\frac{q^2}{\beta^{2}a^{2}}\big]\,
\alpha_{(q,m)}-\alpha_{(q,m)}^{3}=0 .
\end{equation}
It can be regarded as a stationary point of the Lagrangian
\begin{equation}\label{lalpha}
\mathcal{L}^{(q,m)}_{\alpha}=\sqrt{\vert\textrm{det}{g_{E}}\vert}\,
\big[\frac{\hbar^{2}}{2}\, g_{E}^{\mu \nu}\p_{\mu}  \alpha\, \p_{\nu}
\alpha-V_{(q,m)}(\alpha)\big] ,
\end{equation}
where the potential
\begin{equation}
V_{(q,m)}(\alpha)=\varepsilon\frac{a_{0}}{2a}\big[\epsilon m^2+\frac{q^2}{\beta^{2}a^{2}}\big]\,
\alpha^{2}-\frac{1}{4}\,\alpha^{4} ,
\end{equation}
has an inverted mexican hat shape for $(\epsilon,\varepsilon)=(1,1)$. When dealing with the field
$\alpha$ one should be careful since this is not a dynamical
field. Its stress energy tensor should not be included into the
5-dimensional Einstein equations, and expressions like
$\mathcal{L}=\mathcal{L}_{g_{E}}+\int d
\mu(q)(\mathcal{L}_{\psi_{q}}+ \mathcal{L}^{(q,m)}_{\alpha})$ seem unjustified. In summary
\begin{theorem} \label{teop}
Let the field $\Psi$  solve the K-G equation (\ref{kos}) in $(P,
\tilde{g})$ where
\[
\tilde{g}=\varepsilon g_0-a^{2} \tilde{\omega}^{2},
\]
and $g_0$ is a Lorentzian metric. Define  the Einstein metric $g_{E}:=\frac{a}{a_0} \, g_0$. Moreover, let  $\alpha_{(q,m)}$ be a stationary point of the action
\[
\sqrt{\vert\textrm{det}\,{g_{E}}\vert}\,
\big[\frac{\hbar^{2}}{2}\, g_{E}^{\mu \nu}\p_{\mu}  \alpha\, \p_{\nu}
\alpha-\varepsilon\frac{a_{0}}{2a}\big[\epsilon m^2+\frac{q^2}{\beta^{2}a^{2}}\big]\,
\alpha^{2}+\frac{1}{4}\,\alpha^{4} \big] ,
\]
and let
\[
\Psi=\int d\mu(q) \,\psi_{q}(x)  \alpha_{(q,m)}
e^{i\frac{q}{\hbar\beta} y} ,
\]
be the Fourier expansion of $\Psi$
along the fiber, then the fields $\psi_{q}$ satisfy the K-G
equation of a particle of mass $m$ and charge $q$ in $M$
\begin{equation} \label{nap}
D^{q \, \mu}D^{(q,m)}_{\mu}\psi_{(q,m)}+m^{2}\psi_{(q,m)}=0 ,
\end{equation}
which couples these particles to the metric\footnote{It might be convenient to define  $g_{(q,m)}= \alpha_{(q,m)}^{2} g_{E}$ and the K-G equation as $D^{q \, \mu}D^{(q,m)}_{\mu}\psi_{(q,m)}+\psi_{q}=0$ so that the mass is absorbed in the coupling metric, and the mass is read from $\alpha$.} $g_{(q,m)}= \alpha_{(q,m)}^{2} g_{E}/m^2$.
\end{theorem}
The square of the so called vacuum expectation value of $\alpha_{(q,m)}$ is $\varepsilon\frac{a_{0}}{a}\big[\epsilon m^2+\frac{q^2}{\beta^{2}a^{2}}\big]$, so $m^2$ in Eq.\ (\ref{nap}) could be deduced from that.

As a final observation note that the functions $\alpha_{(q,m)}$
are not completely fixed by the differential equation. In fact
there is the freedom related to the choice of initial conditions.
Therefore, for $\hbar \ne 0$ the conformal factor is
considerably constrained but not completely determined. For $\hbar=0$, instead, the conformal factor is completely fixed.

%
%
%

\section{Conclusions}
We have studied the Lorentz force equation in the context of
Kaluza-Klein theory showing  the dependence of proper time on the
scale factor $a$ and on the particle charge-to-mass ratio.
We have  also  given other arguments related to the
solution of the conformal ambiguity problem through Theor.\
\ref{teop}. We have studied in detail the reduction, even with
variable scalar field $a$, of the Klein-Gordon equation showing
that  each Fourier mode is itself a solution of the Klein-Gordon
equation in $M$ where the coupling metric coincides in the limit
$\hbar \to 0$  with the one obtained from the study of the Lorentz
force equation.

The conclusion we draw is that although the Kaluza-Klein theory
still suffers of many fundamental problems one of its necessary
consequences seems to be the dependence of the proper time of charged
particles on the mass, the charge and the scalar field $a$.



\def\cprime{$'$}

\end{document}